\documentstyle[psfig]{mn}

\newif\ifAMStwofonts

\def\la{\mathrel{\hbox{\rlap{\hbox{\lower4pt\hbox{$\sim$}}}\hbox{$<$}}}}
\def\ga{\mathrel{\hbox{\rlap{\hbox{\lower4pt\hbox{$\sim$}}}\hbox{$>$}}}}

\newcommand{\be}{\begin{eqnarray}}
\newcommand{\ee}{\end{eqnarray}}

\newcommand{\msol}{\ifmmode{{\rm M}_\odot}\else{M$_\odot$}\fi}
\newcommand{\foe}{\ifmmode{10^{51}}\else{$10^{51}$}\fi}

\newcommand{\xni}{\ifmmode{{\rm X}_{\rm Ni}}\else{X$_{\rm Ni}$}\fi}
\def\ang{\hbox{\AA}}
\def\Teff{\ifmmode{T_{\rm eff}}\else{\hbox{$T_{\rm eff}$} }\fi}
\def\Rzero{\ifmmode{R_0}\else{\hbox{$R_0$} }\fi}

\def\SP2{{\tt IBM SP2}}

\def\PC2{{\tt PC$^2$}}

\def\logg{\log(g)}

\def\inu{\ifmmode{I_{\nu}}\else{\hbox{$I_{\nu}$} }\fi}
\def\snu{\ifmmode{S_{\nu}}\else{\hbox{$S_{\nu}$} }\fi}
\def\jnu{\ifmmode{J_{\nu}}\else{\hbox{$J_{\nu}$} }\fi}

\def\fep{\ifmmode{{\rm Fe II}}\else\hbox{Fe~II }\fi}

  at 12.0 true pt 

\def\phoenix{{\tt PHOENIX}}

\def\phoenix{{\tt PHOENIX}}

\def\b{\beta}

\def\rout{\ifmmode{r_{\rm out}}\else\hbox{$r_{\rm out}$}\fi}
\def\tmax{\ifmmode{\tau_{\rm max}}\else\hbox{$\tau_{\rm max}$}\fi}
\def\tstd{\ifmmode{\tau_{\rm std}}\else\hbox{$\tau_{\rm std}$}\fi}
\def\vmax{\ifmmode{v_{\rm max}}\else\hbox{$v_{\rm max}$}\fi}
\def\muE{\ifmmode{\mu_{\rm E}}\else\hbox{$\mu_{\rm E}$}\fi} 
\def\pE{\ifmmode{p_{\rm E}}\else\hbox{$p_{\rm E}$}\fi} 
\def\bmax{\ifmmode{\b_{\rm max}}\else\hbox{$\b_{\rm max}$}\fi}
\def\kms{\hbox{$\,$km$\,$s$^{-1}$}}

\def\ang{\hbox{\AA}}

\def\Teff{\hbox{$\,T_{\rm eff}$} }

\def\alog#1{\times 10^{#1}}

\def\rout{\hbox{$r_{\rm out}$} }

\def\chistd{\ifmmode{\chi_{\rm std}}\else\hbox{$\chi_{\rm std}$}\fi}

\def\msol{$M_\odot$}
\def\foe{10^{51}}

\def\xni{{\rm X}_{\rm Ni}}

\def\lstar{\ifmmode{\Lambda^*}\else\hbox{$\Lambda^*$}\fi} 
\def\Rop{\ifmmode{[R_{ij}]}\else\hbox{$[R_{ij}]$}\fi}

\def\Rji{\ifmmode{[R_{ji}]}\else\hbox{$[R_{ji}]$}\fi}
\def\Rstar{\ifmmode{[R_{ij}^*]}\else\hbox{$[R_{ij}^*]$}\fi}

\def\Rjistar{\ifmmode{[R_{ji}^*]}\else\hbox{$[R_{ji}^*]$}\fi}
\def\DRji{\ifmmode{[\Delta R_{ji}]}\else\hbox{$[\Delta R_{ji}]$}\fi}
\def\DRij{\ifmmode{[\Delta R_{ij}]}\else\hbox{$[\Delta R_{ij}]$}\fi}

\def\ns{\ifmmode{N_{\rm s}}          
        \else\hbox{$N_{\rm s}$}\fi}

\def\mat#1{{\bf #1}}     
\def\vek#1{{#1}}         

\newcount\eqcount
\def
  \nummer{
    \global\advance\eqcount by 1
    (\the\eqcount)
  }

\def
  \numadv{
    \global\advance\eqcount by 1
  }

\def
   \numout#1{
     (\the\eqcount #1)
  }

\def\ivek#1#2{\ifmmode{\vek{I}^{#1}_{#2}}
        \else\hbox{$\vek{I}^{#1}_{#2}$}\fi}

\def\tmat#1#2{\ifmmode{\mat{t}^{#1}_{#2}}
        \else\hbox{$\mat{t}^{#1}_{#2}$}\fi}
\def\rmat#1#2{\ifmmode{\mat{r}^{#1}_{#2}}
        \else\hbox{$\mat{r}^{#1}_{#2}$}\fi}
\def\bvek#1#2{\ifmmode{\beta^{#1}_{#2}}
        \else\hbox{$\beta^{#1}_{#2}$}\fi}

\def\lp{\ifmmode{\lambda^+_\tau}           
        \else\hbox{$\lambda^+_\tau$}\fi}
\def\lm{\ifmmode\lambda^-_\tau             
        \else\hbox{$\lambda^-_\tau$}\fi}

\ifoldfss
  \ifCUPmtlplainloaded \else
    \NewTextAlphabet{textbfit} {cmbxti10} {}
    \NewTextAlphabet{textbfss} {cmssbx10} {}
    \NewMathAlphabet{mathbfit} {cmbxti10} {} 
    \NewMathAlphabet{mathbfss} {cmssbx10} {} 
  \fi
  \ifAMStwofonts
    \ifCUPmtlplainloaded \else
      \NewSymbolFont{upmath} {eurm10}
      \NewSymbolFont{AMSa} {msam10}
      \NewMathSymbol{\upi}     {0}{upmath}{19}
      \NewMathSymbol{\umu}     {0}{upmath}{16}
      \NewMathSymbol{\upartial}{0}{upmath}{40}
      \NewMathSymbol{\leqslant}{3}{AMSa}{36}
      \NewMathSymbol{\geqslant}{3}{AMSa}{3E}

    \fi
  \fi
\fi 

\ifnfssone
  \newmathalphabet{\mathit}
  \addtoversion{normal}{\mathit}{cmr}{m}{it}
  \addtoversion{bold}{\mathit}{cmr}{bx}{it}
  \newmathalphabet{\mathbfit} 
  \addtoversion{normal}{\mathbfit}{cmr}{bx}{it}
  \addtoversion{bold}{\mathbfit}{cmr}{bx}{it}
  \newmathalphabet{\mathbfss} 
  \addtoversion{normal}{\mathbfss}{cmss}{bx}{n}
  \addtoversion{bold}{\mathbfss}{cmss}{bx}{n}
  \ifAMStwofonts
    \ifCUPmtlplainloaded \else
      \UseAMStwoboldmath
      \makeatletter
      \new@mathgroup\upmath@group
      \define@mathgroup\mv@normal\upmath@group{eur}{m}{n}
      \define@mathgroup\mv@bold\upmath@group{eur}{b}{n}
      \edef\UPM{\hexnumber\upmath@group}
      \new@mathgroup\amsa@group
      \define@mathgroup\mv@normal\amsa@group{msa}{m}{n}
      \define@mathgroup\mv@bold\amsa@group{msa}{m}{n}
      \edef\AMSa{\hexnumber\amsa@group}
      \makeatother
      \mathchardef\upi="0\UPM19
      \mathchardef\umu="0\UPM16
      \mathchardef\upartial="0\UPM40
      \mathchardef\leqslant="3\AMSa36
      \mathchardef\geqslant="3\AMSa3E
    \fi
  \fi
\fi 

\ifnfsstwo
  \DeclareMathAlphabet{\mathbfit}{OT1}{cmr}{bx}{it}
  \SetMathAlphabet\mathbfit{bold}{OT1}{cmr}{bx}{it}
  \DeclareMathAlphabet{\mathbfss}{OT1}{cmss}{bx}{n}
  \SetMathAlphabet\mathbfss{bold}{OT1}{cmss}{bx}{n}
  \ifAMStwofonts
    \ifCUPmtlplainloaded \else
      \DeclareSymbolFont{UPM}{U}{eur}{m}{n}
      \SetSymbolFont{UPM}{bold}{U}{eur}{b}{n}
      \DeclareSymbolFont{AMSa}{U}{msa}{m}{n}
      \DeclareMathSymbol{\upi}{0}{UPM}{"19}
      \DeclareMathSymbol{\umu}{0}{UPM}{"16}
      \DeclareMathSymbol{\upartial}{0}{UPM}{"40}
      \DeclareMathSymbol{\leqslant}{3}{AMSa}{"36}
      \DeclareMathSymbol{\geqslant}{3}{AMSa}{"3E}
    \fi
  \fi
\fi 

\ifCUPmtlplainloaded \else
  \ifAMStwofonts \else 
    \def\upi{\pi}
    \def\umu{\mu}
    \def\upartial{\partial}
  \fi
\fi

\title[Pistinner et al.:   Low Metallicity Giant H~II Regions ]
{ Spectroscopy Of Low Metallicity 
Giant H~II Regions: \\ A grid of low Metallicity Stellar atmospheres}
\author[S.L. Pistinner, P. H. Hauschildt, D. Eichler, \& E. A. Baron ]
{S.L. Pistinner$^{1}$, P. H. Hauschildt$^{2}$, D. Eichler$^{3}$ \& 
E. A. Baron$^{4}$ \\ 
$^{1}$Dept. of Applied Mathematics
Israel Institute for Biological Research \\
P.O.B 19 Nes-Ziona 74100 Israel \\
$^{2}$Dept. of Physics \& Astronomy, 
The University of Georgia,
Athens, GA 30602-2451, USA \\
yeti@hal.physast.uga.edu \\
$^{3}$Department of  Physics, Ben-Gurion University 
Beer-Sheba, 84105, Israel \\
eichler@bguvms.bgu.ac.il \\
$^{4}$Dept. of Physics \& Astronomy, 
University of Oklahoma,
Norman OK 73019-0225, USA \\
baron@mail.nhn.ou.edu }
\date{Accepted xx. Received xx}

\pubyear{1998}

\begin{document}
\bibliographystyle{HII} 

\maketitle


\begin{abstract}
We calculate a grid of spherically symmetric OB stellar atmospheres at low
metallicities, including both non-local thermodynamic equilibrium (NLTE) and
metal line blanketing effects. This is done to assess the uncertainties in
helium abundance determination by nebular codes due to input stellar atmosphere
models. The more sophisticated stellar atmosphere models we use can differ from
LTE models by as much as 40 percent in the ratio of He to H-ionizing photons.
\end{abstract}


\section{INTRODUCTION} \label{sec:introduction}
Giant extra-galactic H~II regions (GEHR) are interesting objects for
many reasons  (cf. Shields 1990 for a review).  However, 
low metallicity HII regions draw the most attention. This subgroup
of low metallicity GEHR is believed to have had little chemical evolution
 over a Hubble time. Therefore, it is widely accepted that
GEHR offer an opportunity to measure the primordial chemical composition
of the universe.

The primordial composition of the universe is predicted by big-bang
nucleosynthesis theory. If low metallicity GEHR provide a way to
determine the primordial element abundance then they allow testing of some
big-bang cosmology predictions, in particular the primordial helium
abundance $Y^{P}$, which has implications for the baryonic fraction
of matter in the universe and may limit the number of exotic light
particle species. Although low metallicity
GEHR show little chemical evolution, some chemical evolution did take
place  over a Hubble time, thereby requiring an extrapolation of the
``measured'' helium abundance $Y^{obs}$ (as a function of the observed
metallicity) to zero metallicity.  This  procedure yields the  value
of $Y^{P}$, the quantity of interest.  The extrapolation techniques to
zero metallicity are discussed by Pegal {et.~al.} (1992).  
In this paper we
concentrate on systematic errors in the determination of $Y^{obs}$ from
low metallicity GEHR introduced by systematic changes of the radiation
field that ionizes the nebula.

The use of nebular spectroscopy to measure element abundance is model
dependent. The internal physics of GEHR is therefore of prime importance. 
The uncertainties include: a) the spectrum of the ionizing 
low metallicity stars,
b) the filamentary structure of GEHR, c)  supersonic turbulence inferred
from radio observations, d)  radiative transfer effects within the nebula
(Terlevich, Skillman \& Terlevich 1995 ;Sasselov \& Goldwirth 1995).

The uncertainty of  $Y^{obs}$ determinations is required to be no more
than several percent
 in order to derive accurate enough values of $Y^P$ for, say, establishing
the number of light neutrino species. Typically the
value of $Y^{obs}$ is derived from recombination lines of singly and
doubly ionized helium \cite{skill94} which are compared to H$_{\alpha}$
emission.  This implies that an essential assumption is that the fraction
of neutral helium in the nebula is negligible.  If the
fraction of
neutral Helium is several percent, then the value of $Y^{obs}$ cannot be
determined
to the required accuracy.  Therefore, the helium ionization degree in
H~II regions has been the subject of numerous meticulous studies. Two
problems are often mentioned in this context: i) the filamentary structure
of the nebula and ii) the uncertainties in the underlying ionizing
(stellar)
radiation field. The latter uncertainties can translate into
uncertainties in relative sizes of the He and H ionization zones.

This paper does not attempt to deal with problems associated with the
filamentary structure of the nebula. However, the existence of this
problem must be noted. It is observationally established that most of the
line emission of GEHR comes from a rather small filamentary volume. The
volume filling factor of these filaments is about 1\%. They move relative
to each other with a velocity dispersion which is supersonic.  Moreover,
radio observations reveal that in about half of the H~II regions in
M51 (not a low metallicity region) the radio emission originates in a
non-thermal component within the filaments. The origin and 
dynamics of the filaments are not fully understood. Clearly, if they
contain any neutral or singly ionized He, then the He/H abundance
determination is hindered by this source of inaccuracy. 
Thus, before any final statement can be made, one
requires strong observational constraints
of the properties of the matter in these filaments or a
reliable theoretical model
to understand the origin and the detailed physics of the filaments.

Due to the complexity of the nebular filamentary structure most
authors have invoked homogeneous nebular models, in slab or spherical
geometry. These spherical and shell homogeneous models constrain the
ionization degree of helium to be 0.98 if the effective temperature of
the ionizing stars is above 38,000K \cite{diner86,skill89}.  However,
these authors note that if the GEHR  is actually an ensemble of much
smaller H~II  regions which are ionized by stars of different effective
temperatures (lower than 38,000K, Dinerstein  1989), filamentary neutral
helium structures might form within the nebula.  Even if this problem
is ignored and one assumes that all ionizing stars have effective
temperatures above 38,000K, there are uncertainties due to a lack of a
state-of-the-art grid of model atmospheres in the required range of
interest \cite{shields90}. Recently, some effort has been put into
this problem by \cite{gar96,gar97}. They attempt to account for the
filamentation by using shell models of the nebula, then, using 
the photo-ionization code CLOUDY \cite{cloudy}, they 
fit the stellar and nebular spectra self-consistently.
They conclude \cite{gar97} that contamination from young evolved
stellar population e.g. 
Wolf-Rayet stars, is of importance and must be taken into
account.

The need for consistent spectral modeling of GEHR has been recognized
by \cite{gar94}, who have used \cite{kur92} LTE line blanketed
models with a range of abundances as input to CLOUDY.  However, they
express their concern that NLTE effects might become important in the
relevant range of parameters.  This lack of state-of-the-art NLTE line
blanketed model stellar atmospheres is troubling, because most of the
ionizing spectra of the nebula is in the UV range, which is strongly
metal line blanketed. Typically for solar abundance stars, photons in
this  spectral region are strongly affected by the presence of many
spectral lines.  Yet, the input stellar model atmospheres used so far
assumed either a pure hydrogen helium mixture in NLTE, or  assumed
LTE for the metal lines. This introduced a source of uncertainty
that this paper addresses.  In addition, \cite{kur94} has improved
and modified the line data set of \cite{kur92}, and these effects are
taken into account as well.

There is already a substantial literature on non-LTE effects (some of
the classical papers are: Auer \& Mihalas 1972, Kudritzki 1973, 1976,
1979, 1988, Husfeld et al. 1984).  Our purpose here is to evaluate their
contribution to the uncertainties in He abundance determination. In
recent years stellar atmosphere and radiative transfer codes have
become very sophisticated in a way which allows much more of the
important physics to be taken into account.  We use such a code to
construct a grid of low-metallicity hot stellar atmosphere models
including NLTE and metal line blanketing.  This allows us to assess
and reduce one of the uncertainties in the determination of $Y^{obs}$.
We provide the spectra of these models in machine readable format at
{\tt http://dilbert.physast.uga.edu/\symbol{'176}yeti}.

The structure of this paper is as follows: in \S~\ref{sec:phoenix} we
discuss briefly the stellar atmosphere code \phoenix\ , and describe in
detail the input physics of stellar spectrum models. We then provide in
\S~\ref{sec:res} some examples of the variation of the spectrum in the
UV range.  We present the ionization parameters of helium and hydrogen,
depending on various assumptions, and assess the variation of these
quantities as functions of the stellar atmosphere model input parameters.
We summarize our conclusions in \S~\ref{sec:con}.

\section{Methods and Models}\label{sec:phoenix} 

In order to investigate the importance of NLTE effects on the formation of OB
star spectra and ionizing photon fluxes, full and very detailed NLTE model
calculations are required in order to model the effects of NLTE on the very
important EUV and UV metal lines. This means that the multi-level NLTE rate
equations must be solved self-consistently and simultaneously with the
radiative transfer and energy equations, including the effects of line
blanketing and the extension of the atmosphere. For the purpose of this
analysis we use our multi-purpose stellar atmosphere code \phoenix.  
\phoenix~[version 9.1~\cite{fe2nova,nova-97,parapap,parapap2}] 
uses a special relativistic
spherical radiative transfer and an equation of state (EOS) including more than
300 atoms and ions (39 elements with up to 26 ionization stages). The
temperature correction is based on a modified (for NLTE and scattering)
Uns\"old-Lucy method that converges quickly and is numerically stable.
See \cite{jcam} for details on the numerical methods used in \phoenix.

Both the NLTE and LTE (background) lines are treated with a direct
opacity sampling method.  However, we do {\em not} use pre-computed
opacity sampling tables. We dynamically select the relevant LTE background
lines from master line lists at the beginning of each iteration and sum up
the contribution of every line to compute the total line opacity at {\em
arbitrary} wavelength points. The latter is crucial in NLTE calculations
in which the wavelength grid is both irregular and variable (from
iteration to iteration due to changes in the physical conditions). This
approach also allows detailed and depth dependent line profiles to be used
during the iterations. To make this method computationally efficient, we
employ modern numerical techniques, e.g., vectorized and parallel block
algorithms with high data locality \cite{parapap}, and we use high-end
workstations or parallel supercomputers for the model calculations.
In the calculations we present in this paper, we have set the statistical
(treated as micro-turbulence) velocity $\xi$ to $2\kms$. We include 
LTE background lines (i.e., lines of species that
are not treated explicitly in NLTE) if they are stronger than a threshold
$\Gamma\equiv \chi_l/\kappa_c=10^{-4}$, where $\chi_l$ is the extinction
coefficient of the line at the line center and $\kappa_c$ is the local
b-f absorption coefficient. This typically leads to about $2\alog{6}$
LTE background lines.  The line profiles of these lines are taken to be
depth-dependent profiles (with Voigt profiles for the strong lines and
Gauss profiles for weak lines). We have verified in test calculations
that the details of the LTE background line profiles and the threshold
$\Gamma$ do not have a significant effect on the model structure and the
synthetic spectra. However, the LTE background lines should be included
in detailed model calculations because their cumulative effect can change
the structure and the synthetic spectra.  In addition, we include about
2000 photo-ionization cross sections for atoms and ions (Verner \&
Yakovlev 1995).

\phoenix\ is a full multi-level NLTE code, i.e., NLTE effects are
considered self-consistently in the model calculations, including the
radiative transfer calculations and the temperature corrections.
Hauschildt \& Baron (1995) have extended the
numerical method developed by Hauschildt (1993) for
NLTE calculations with a very detailed model atom of Fe~II. In the
calculations presented in this paper, we use a significantly enlarged
set of NLTE species, namely H, He~I--II, Mg~II, Ca~II, Ne~I, C~I--IV,
N~I--VI, O~I--VI, S~II--III, and Si~II--III, \cite[for a complete list of NLTE
species available in \phoenix\ 9.1 see]{nova-97}.  Here, we do
{\em not} use the NLTE treatment for Li~I, Na~I, Fe~I--III, Co~I---III
and Ti~I--III because these ionization stages are not important in
metal-poor OB star atmospheres and particularly Fe, Co, and Ti NLTE would
considerably increase the CPU time for the model calculations with little
additional improvements in the results.  We include a total of  2120 NLTE
levels and 14,080 NLTE primary lines in the calculations presented here,
nearly a factor of 5 more levels and lines
than in our previous calculations \cite{fe2nova}. The construction of
the model atoms is described in \cite{nova-97} and references therein.

\subsection{NLTE Calculation Method}

The large number of transitions that have to be included in realistic
models of the NLTE line formation require an efficient method for the
numerical solution of the multi-level NLTE radiative transfer and
model calculation  problem.  The model atoms
used here include more than 14,080 individual NLTE lines plus a large
number of weak background transitions. Classical techniques, such as
the complete linearization or the Equivalent Two Level Atom methods,
are computationally prohibitive. \phoenix\ performs its calculations
in the co-moving frame for expanding atmospheres
(e.g., stellar winds, novae and supernovae), therefore, approaches
such as Anderson's multi-group scheme \cite{and87,anderson89} or
extensions of the opacity distribution function method \cite{HubLan95}
cannot be applied. Simple approximations such as the Sobolev
method, are very inaccurate in problems in which lines overlap strongly
and make a significant continuum contribution (important for weak lines),
as is the case for nova (and SN) atmospheres cf. \cite{fe2nova,fe2sn}.

We use the multi-level rate-operator splitting (or preconditioning)
method described by \cite{casspap}. This method solves the non-grey,
spherically symmetric, special relativistic equation of radiative
transfer in the co-moving (Lagrangian) frame using the operator splitting
method described in \cite{s3pap}.  Details of the method are described
in \cite{fe2pap}, so we do not repeat the detailed description here.
For all primary NLTE lines the radiative rates and the approximate rate
operators \cite{casspap} are computed and included in the iteration
process.  Secondary NLTE lines are included as background transitions
for completeness but are insignificant for the model structure and the
synthetic spectra (the model atoms have been explicitly constructed so
that all important lines are primary lines).

\subsection{Atmosphere Models}
\begin{figure}
\centerline{\psfig{file=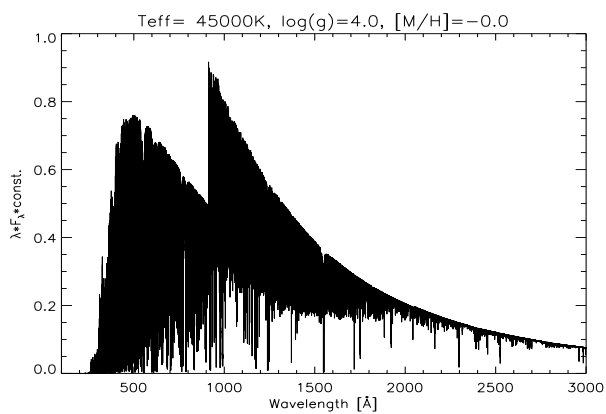,height=5.5cm,width=8.50cm,angle=90}}
\caption{\label{hr} The 
NLTE spectrum for $\Teff=45,000\,$K, $\logg=4.0$ 
and solar metallicity at the nominal
resolution used in the model calculations.} 
\end{figure}

We have computed a small grid of spherically symmetric {\em static} models to
investigate the effects of NLTE on the structure and the spectra of metal poor
OB-stars.  The models include the NLTE treatment as discussed above as well as
the standard \phoenix\ NLTE generalized equation of state and additional LTE
background lines (about 2 million atomic lines).  Aufdenberg et. al. 
(1998) showed that the
{\em combined} effects of line blanketing and spherical geometry can
dramatically change the short wavelength spectrum of the models, therefore, we
calculated all models presented here using spherical geometry (incl.\
spherically symmetric radiative transfer). Models that include the effects of
a stellar wind on the structure of the atmosphere and the emitted spectrum
are currently being calculated (Aufdenberg et al, in preparation).
The NLTE effects are included in
both the temperature iterations (so that the structure of the models includes
NLTE effects) and all radiative transfer calculations.  For each primary NLTE
line we add 3 to 5 wavelength points to the overall wavelength grid.  This
procedure typically leads to about 55,000--150,000 wavelength points for the
model computations and the synthetic spectrum calculations. Test calculations
have shown that the resulting overall wavelength grid is completely adequate 
for NLTE calculations in static and expanding media \cite{fe2pap}. We have also
calculated LTE continuum and line blanketed models for comparison.  In figure
\ref{hr} we show a synthetic spectrum at the nominal resolution, illustrating
the effect of low resolution  on the appearance of the spectrum. 

\begin{figure}
\centerline{\psfig{file=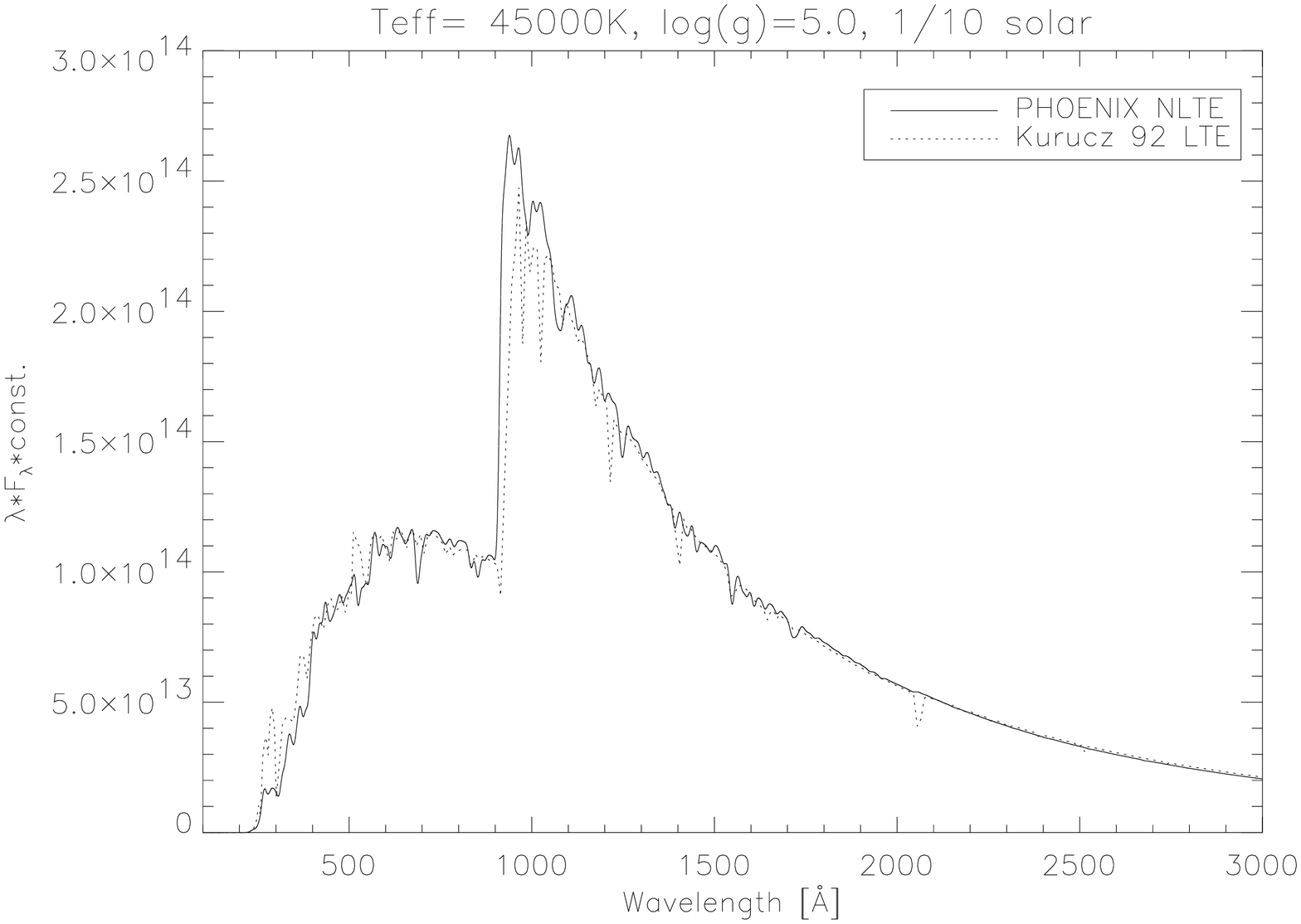,height=5.5cm,width=8.50cm,angle=90}}
\caption{\label{comp1} Comparison between a 
\phoenix\ NLTE spectrum and a Kurucz 1992
LTE spectrum for $\Teff=45,000\,$K, $\logg=5.0$ 
and 1/10 solar metallicity. The
\phoenix\ sspectrum has been degraded in 
resolution by convolving it with a Gaussian
kernel with $6\ang$ half-width.}
\end{figure}

\begin{figure}
\centerline{\psfig{file=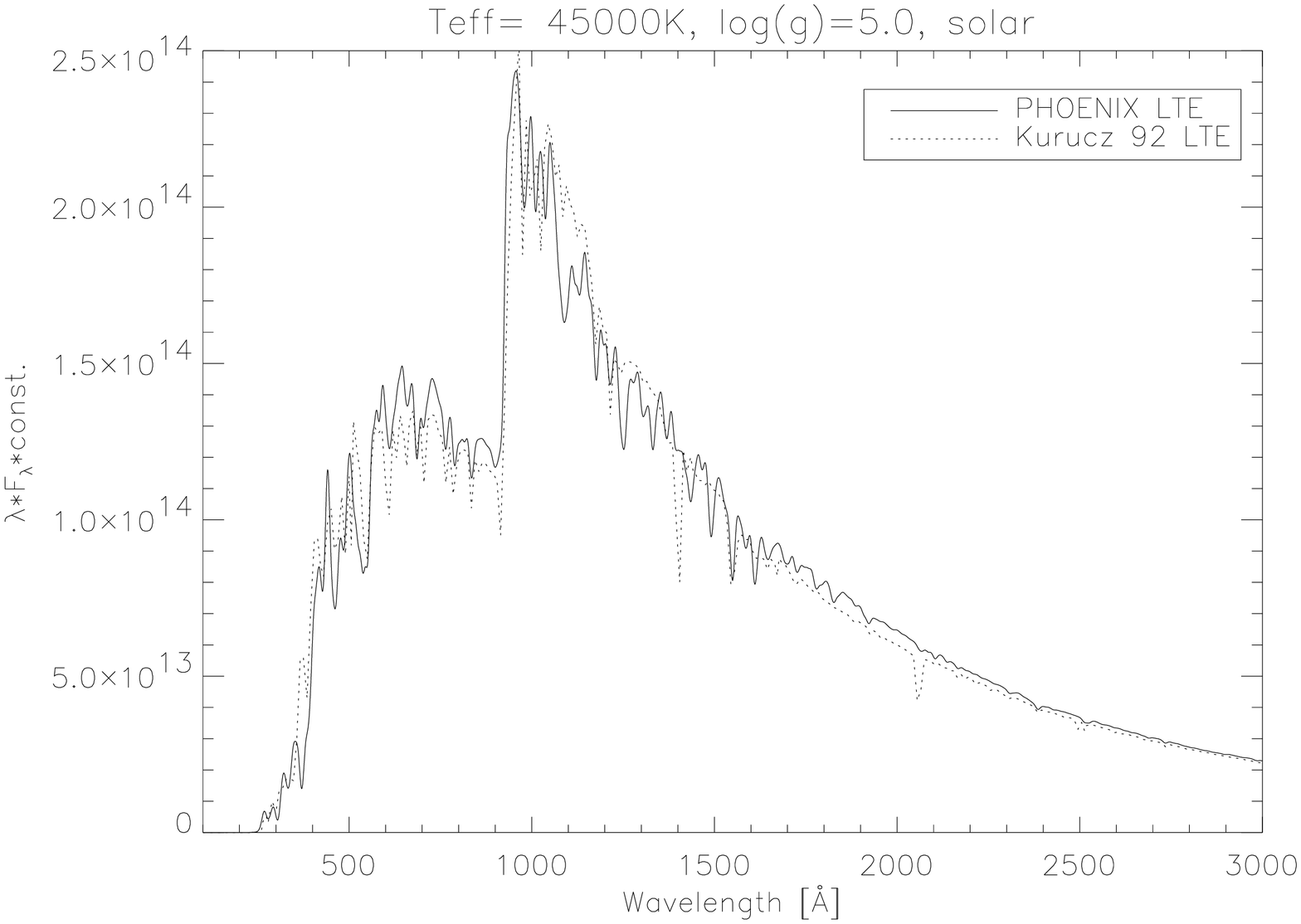,height=5.5cm,width=8.50cm,angle=90}}
\caption{\label{comp2} Comparison between a 
\phoenix\ NLTE spectrum and a Kurucz 1992
LTE spectrum for $\Teff=45,000\,$K, $\logg=5.0$ and solar metallicity. The
\phoenix\ sspectrum has been degraded 
in resolution by convolving it with a Gaussian
kernel with $6\ang$ half-width.}
\end{figure}

We compare our LTE and NLTE models to the Kurucz~92 LTE set of synthetic
spectra in Figs.~\ref{comp1} and \ref{comp2}. The \phoenix\ spectra have
been convolved with a Gaussian kernel of $6\ang$ half-width to make the
resolutions of the different sets of spectra comparable.  In general
the spectra are very similar. We use an updated version of the Kurucz
atomic line lists; this could explain the differences between the LTE
models. In addition, the \phoenix\ models allow for line scattering and
NLTE (Fig.~\ref{comp2}) whereas the Kurucz models are calculated using
complete LTE (no line scattering). 

\section{Results}\label{sec:res}
All model atmospheres presented in this paper have $\log(g)=4.0$.
Three effective temperatures (A, B, and C respectively)  $T_{eff}=38,000$,
45,000, and $55,000\,$K,
have been considered. For each of these $\Teff$ we have computed models
with 10\%, 5\% and 2\% solar metal abundances(1-3, 4-6, 7-9 respectively).
In each case NLTE and LTE
line blanketed models as well as LTE continuum models were calculated. The
models and the resulting ionization parameters $q_{0}$: the base 
ten-logarithm of the
emission rate of HI Lyman continuum  photons
\begin{equation}
q_0=\log_{10}\left(4 \pi^{2} \int_{0}^{\lambda_{L\alpha}}
d \lambda \frac{F_{\lambda} \lambda }{hc}\right),
\end{equation}
and $q_{1}$ (helium continuum) are presented in table 1. In addition 
we have computed for each temperature a line blanketed 
LTE solar metallicity model (AS, BS, CS), to 
assess quantitative differences which could result from a contaminating population of
of young evolved
stars that may be present in the HII region.

We start by considering low metallicity models.  As expected,
the resulting ionization photon fluxes are most sensitive to the
temperature. 
The dependence in $q_{1}$ on the metallicity
is typically  weak. The strongest variation is among the
different model type themselves (LTE vs. NLTE). 
For the higher temperatures the differences in  $q_0$ and $q_1$ are typically
of order 0.02 or less. For models A1-9, the differences in $q_1$ between LTE
and NLTE models can be as large as 0.19, meaning that the amount of 
He ionizing flux differs by a factor of $10^{0.19}$. As the differences in 
$q_{o}$ is only about 0.02, the implied difference is the ratio
of helium-ionizing flux to hydrogen-ionizing flux is about 
$10^{0.17}\sim 1.45$.

\begin{figure}
\centerline{\psfig{file=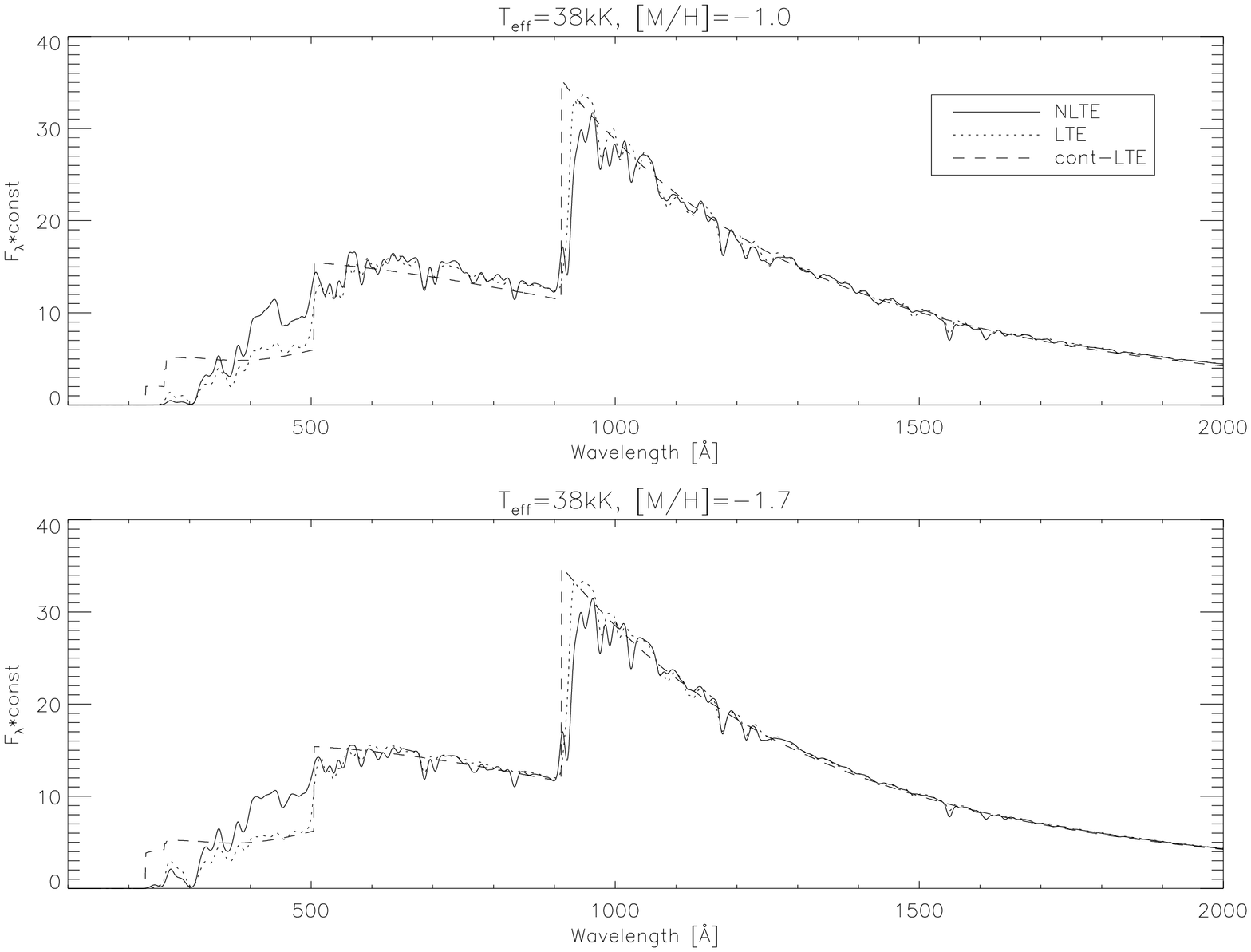,height=5.5cm,width=8.50cm,angle=90}}
\caption{\label{tim1}
Models with $T_{eff}=38$kK, Continuum LTE, Line LTE and NLTE models are 
overlaid. Top panel 10\% solar metal abundance, 
bottom panel 2\% solar metal abundance.}
\end{figure}

\begin{figure}
  \centerline{\psfig{file=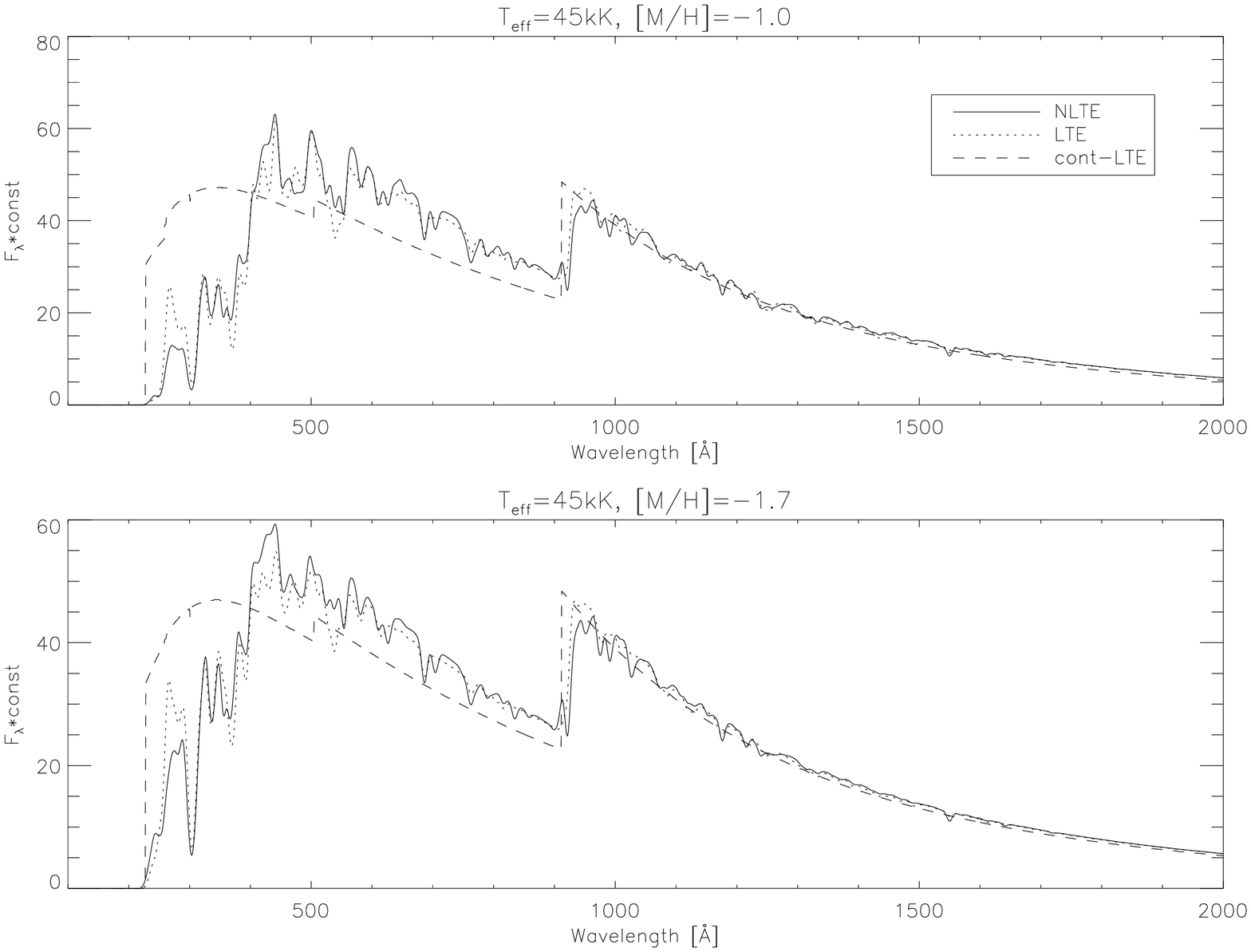,height=5.5cm,width=8.50cm,angle=90}}
\caption{\label{tim2}
Models with $T_{eff}=45$kK, Continuum LTE, Line LTE and NLTE models are 
overlaid. Top panel 10\% solar metal abundance, 
bottom panel 2\% solar metal abundance.}
\end{figure}

\begin{figure}
  \centerline{\psfig{file=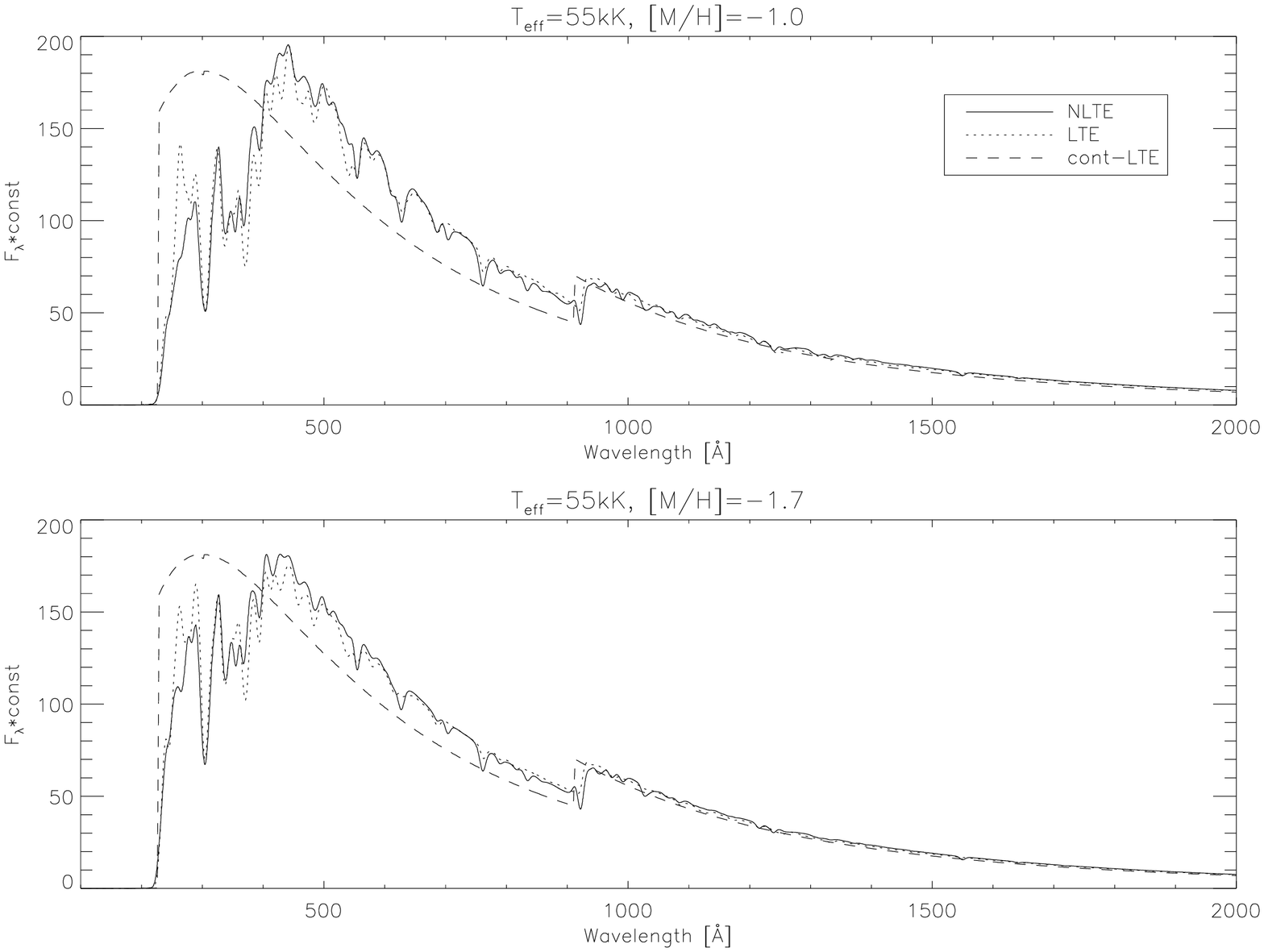,height=5.5cm,width=8.50cm,angle=90}}
\caption{\label{tim3}
Models with $T_{eff}=55$kK, Continuum LTE, Line LTE and NLTE models are 
overlaid. 
Top panel 10\% solar metal abundance, bottom panel 2\% solar metal abundance.}
\end{figure}

We compare the spectra of the NLTE, LTE line 
blanketed and LTE continuum models for three effective temperatures
in Figs.~\ref{tim1} to
\ref{tim3}.  The effects of line blanketing on the overall shape of
the spectra is significant, even at the lowest metallicity that we have
considered. The differences between LTE and NLTE spectra are smaller
than the effects of line blanketing by itself. 
Longward of 1000 \AA, NLTE effects on the spectra are small,
especially near the threshold temperature of 
$\Teff=38,000\,$ and would thus have small if any impact on 
near-UV observations.

\begin{figure}
\centerline{\psfig{file=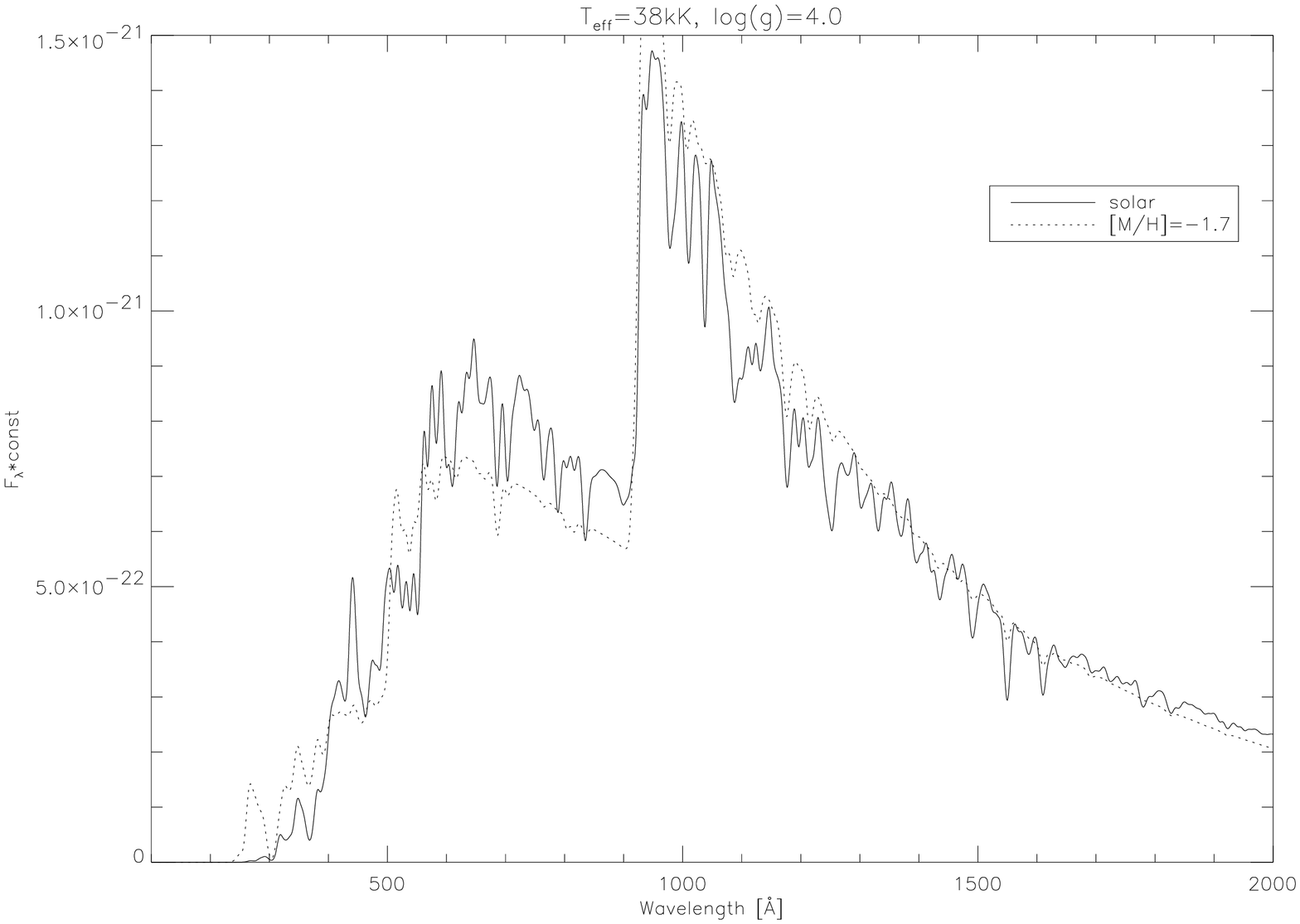,height=5.5cm,width=8.50cm,angle=90}}
\caption{\label{mh1} Comparison between 
Synthetic spectra for $T_{eff}=38$kK for solar and 2\%  solar metal abundances.}
\end{figure}

\begin{figure}
\centerline{\psfig{file=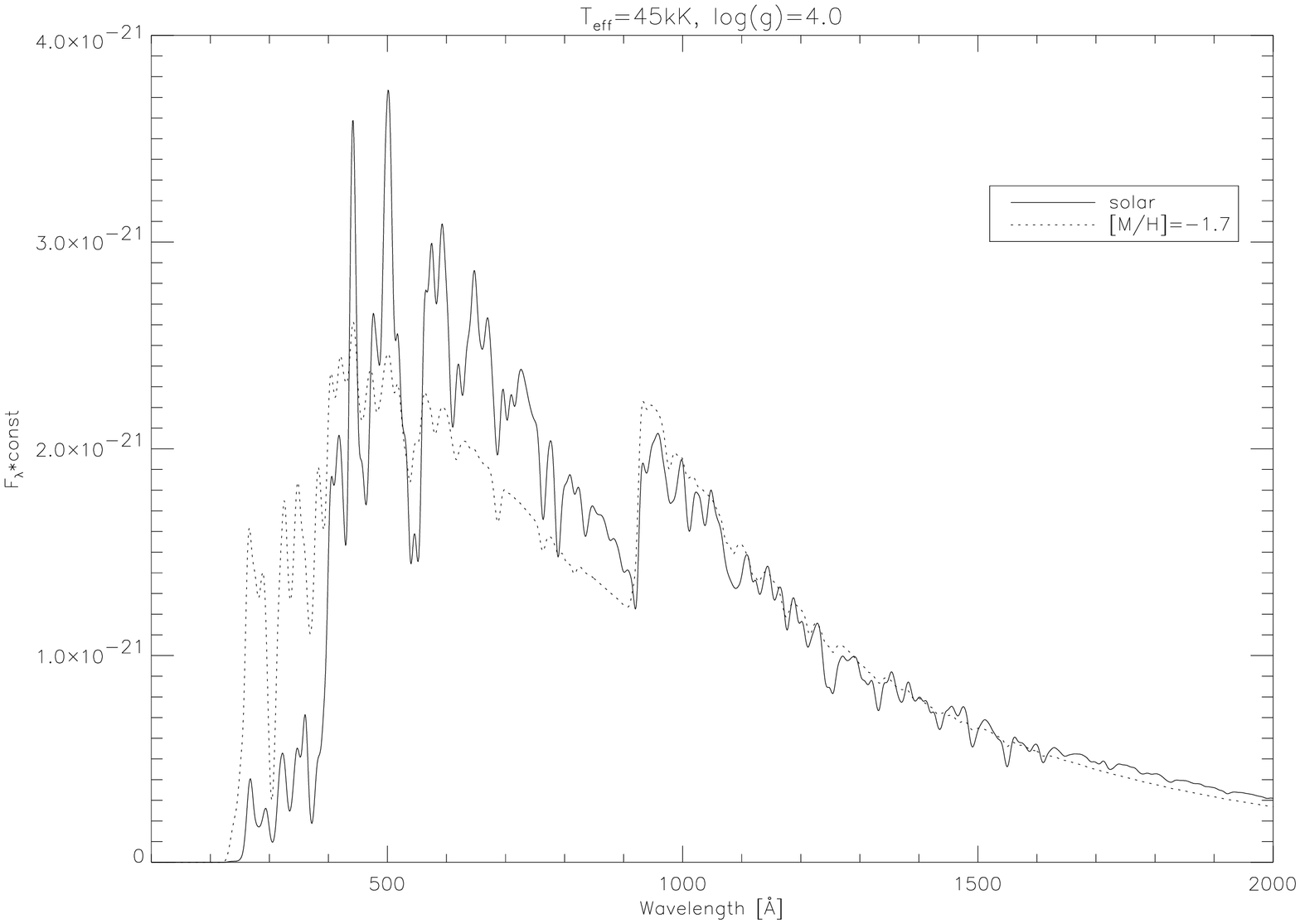,height=5.5cm,width=8.50cm,angle=90}}
\caption{\label{mh2} Comparison between
Synthetic spectra for $T_{eff}=45$kK for solar and 2\%  solar metal abundances.}
\end{figure}

\begin{figure}
\centerline{\psfig{file=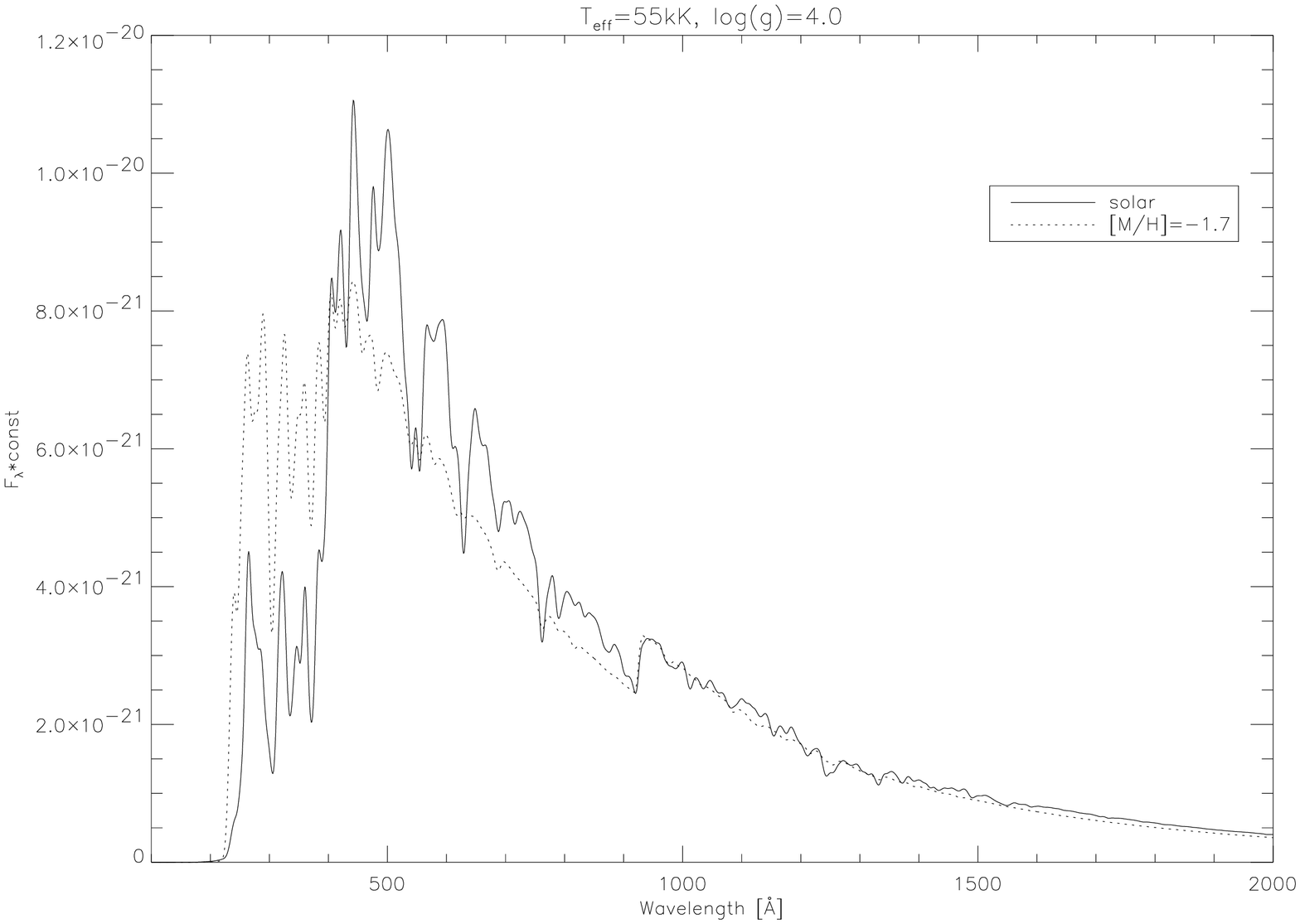,height=5.5cm,width=8.50cm,angle=90}}
\caption{\label{mh3} Comparison between
Synthetic spectra for $T_{eff}=55$kK for solar and 2\%  solar metal abundances.}
\end{figure}

In figures \ref{mh1} to \ref{mh3} we show comparisons between
solar metallicity spectra and spectra computed for 1/10 of the solar
metallicity. In the solar metallicity spectra the metal lines are always
stronger  than in the spectra for 1/10 solar metallicity
(though  the Lyman edge is typically not quite as strong as in the 1/10
solar metallicity spectra).  The differences in the value of $q_{1}$
are  11\%.  This implies that if 10\% of the stars are younger
higher metallicity stars, the differences in the total ionizing number of
photons exciting
the surrounding nebula would be about 1\%. However, the differences
between the spectra are considerably more pronounced. The resulting
effects from this fact have been studied recently by \cite{gar97},
who found that the Wolf-Rayet contribution typically dominates this
effect.

\section{Conclusions}\label{sec:con}
Our synthetic spectra show that the predicted degree of helium
ionization in GEHR could vary significantly depending on
different model assumptions about the stellar atmospheres. 
In the end, the computed value of [He]/[H] in the nebula depends on the
fraction of ionization contributed by the ``relatively cool'' hot stars,
as well as the geometry of the HII region. As the number of
recombinations in an HII region equals the number of ionization, one
might expect that the depths of the ionization zones for He and H would
scale roughly as the cube root of the ionizing fluxes (only roughly
because He-ionizing radiation can also ionize H).
On the other hand, the uncertainties discussed in this paper, i.e. those
due to the assumption of a {\em static} atmosphere,
might be augmented by additional uncertainties in line blanketing
effects, such as the existence of winds 
(Gabler et al., 1989, 1991, 1992; Najarro et al. 1996), 
shocks, etc., which are beyond the
scope of this paper.
The existence of additional uncertainties in the nebular dynamics, such as
those implied by the presence of filaments, also need to be investigated
farther.
The extent to which the theoretical uncertainty can be gauged by scatter
in the data is also hard to assess here, because much of it may be
systematic.

 \smallskip 
\begin{small} 

\noindent{\em Acknowledgments:}
Part of this work was carried while S.P. was a visiting Minerva 
fellow at the Max-Planck Institute f\"ur Astrophysik.  
The work was supported in part by US-Israel 
BSF grant 1802504, by NASA ATP grant NAG 5-3018 and LTSA grant 
NAG 5-3619 and by NSF grant AST-9720704 to UGA, and by NSF grant
AST-9417242, NASA grant NAG5-3505 and an IBM SUR grant to the University
of Oklahoma.  Some of the calculations presented in this paper were
performed on the IBM SP2 of the UGA UCNS, at the San Diego Supercomputer
Center (SDSC) and the Cornell Theory Center (CTC), with support from
the National Science Foundation. We thank all these institutions for a
generous allocation of computer time. 
\end{small}

\clearpage

\begin{table}
\caption[]{\label{modtab} Models Input Parameters and Results. Cont stands 
for continuum LTE models, LTE for line blanketed LTE model and NLTE for
NLTE line blanketed models}
\smallskip
\begin{tabular}{*{6}{l}}
\hline
\hline
 Model &Type & $T_{eff}$& $\lbrack M/H\rbrack$ &$q_{0}$&$ q_{1}$ \\
\hline
A1      &NLTE& 38,000K  & -1. &                  24.0443 & 23.1560  \\
A2      &LTE~& 38,000K  & -1. &                  24.0232 & 22.9831  \\
A3      &Cont& 38,000K  & -1. &                  24.0226 & 23.0756  \\
A4      &NLTE& 38,000K  & -1.3&                  24.0351 & 23.1610  \\
A5      &LTE~& 38,000K  & -1.3&                  24.0204 & 22.9928  \\
A6      &Cont& 38,000K  & -1.3&                  24.0307 & 23.0976  \\
A7      &NLTE& 38,000K  & -1.7&                  24.0311 & 23.1885  \\
A8      &LTE~& 38,000K  & -1.7&                  24.0164 & 22.9959  \\
A9      &Cont& 38,000K  & -1.7&                  24.0292 & 23.0991  \\
\hline
AS      &LTE~& 38,000K  &  0  &                  24.0640 & 22.9930  \\
\hline
B1      &NLTE& 45,000K  & -1. &                  24.5376 & 23.8949  \\
B2      &LTE~& 45,000K  & -1. &                  24.5345 & 23.9063  \\
B3      &Cont& 45,000K  & -1. &                  24.5113 & 24.0245  \\
B4      &NLTE& 45,000K  & -1.3&                  24.5322 & 23.9171  \\
B5      &LTE~& 45,000K  & -1.3&                  24.5345 & 23.9063  \\
B6      &Cont& 45,000K  & -1.3&                  24.5100 & 24.0246  \\
B7      &NLTE& 45,000K  & -1.7&                  24.5282 & 23.9497  \\
B8      &LTE~& 45,000K  & -1.7&                  24.5247 & 23.9512  \\
B9      &Cont& 45,000K  & -1.7&                  24.5092 & 24.0246  \\
\hline
BS      &LTE~& 45,000K  &   0 &                  24.5620 & 23.818   \\
\hline                  
C1      &NLTE& 55,000K  & -1. &                  24.9812 & 24.5081  \\
C2      &LTE~& 55,000K  & -1. &                  24.9842 & 24.5065  \\
C3      &Cont& 55,000K  & -1. &                  24.9525 & 24.5860  \\
C4      &NLTE& 55,000K  & -1.3&                  24.9766 & 24.5197  \\
C5      &LTE~& 55,000K  & -1.3&                  24.9789 & 24.5196  \\
C6      &Cont& 55,000K  & -1.3&                  24.9521 & 24.5858  \\
C7      &NLTE& 55,000K  & -1.7&                  24.9730 & 24.5389  \\
C8      &LTE~& 55,000K  & -1.7&                  24.9734 & 24.5342  \\
C9      &Cont& 55,000K  & -1.7&                  24.9521 & 24.5859  \\
\hline
CS      &LTE~& 55,000K &   0  &                   25.0030 & 24.4670 \\
\hline
\hline
\end{tabular}
\end{table}

\clearpage
\section{Figures}

\end{document}
\end